\documentclass[aps,preprint,nofootinbib]{revtex4-1}
\usepackage{graphicx}
\usepackage{amsmath,amssymb}
\usepackage{url}
\usepackage{color}
\usepackage{epstopdf}
\newcommand{\lsim}{\mathrel{\mathop{\kern 0pt \rlap
  {\raise.2ex\hbox{$<$}}}
  \lower.9ex\hbox{\kern-.190em $\sim$}}}
\newcommand{\gsim}{\mathrel{\mathop{\kern 0pt \rlap
  {\raise.2ex\hbox{$>$}}}
  \lower.9ex\hbox{\kern-.190em $\sim$}}}

\begin{document}
\title{The Top Window for dark matter}
\author{Kingman Cheung$^{1,2,3}$, Kentarou Mawatari$^{4,5}$, \\
Eibun Senaha$^{3}$, Po-Yan Tseng$^{2}$, and Tzu-Chiang Yuan$^6$}

\affiliation{
$^1$Division of Quantum Phases \& Devices, School of Physics, 
Konkuk university, Seoul 143-701, Korea \\
$^2$Department of Physics, National Tsing Hua University, 
Hsinchu 300, Taiwan
\\
$^3$Physics Division, National Center for Theoretical Sciences,
Hsinchu 300, Taiwan
\\
$^4$Institut f\"{u}r Theoretische Physik, Universit\"{a}t Heidelberg, 
Philosophenweg 16, D-69120 Heidelberg, Germany
\\
$^5$Theoretische Natuurkunde and IIHE/ELEM, Vrije Universiteit Brussel,
and International Solvay Institutes,
Pleinlaan 2, B-1050 Brussels, Belgium 
\footnote{Address since October 2010}
\\
$^6$Institute of Physics, Academia Sinica, Nankang, Taipei 11529, Taiwan
}

\date{\today}

\begin{abstract}
We investigate a scenario that the top quark is the only window
to the dark matter particle.  We use the effective Lagrangian
approach to write down the interaction between the top quark and the
dark matter particle.  Requiring the dark matter satisfying the
relic density we obtain the size of the effective interaction.  
We show that the scenario can be made consistent with the direct and indirect
detection experiments by adjusting the size of the effective coupling.  
Finally, we calculate the production cross section
for $t\bar t + \chi \bar \chi$ at the 
Large Hadron Collider (LHC), which will give rise to
an interesting signature of a top-pair plus large missing energy.
\end{abstract}
\maketitle

\section{Introduction}
The presence of cold dark matter (CDM) in our Universe is now well established
by a number of observational experiments, especially the very precise 
measurement of the cosmic microwave background radiation
in the Wilkinson Microwave Anisotropy Probe (WMAP) experiment \cite{wmap}.
The measured value of the CDM relic density is
\[
 \Omega_{\rm CDM}\, h^2 = 0.1099 \;\pm 0.0062 \;,
\]
where $h$ is the Hubble constant in units of $100$ km/Mpc/s.
Though the gravitation nature of the dark matter is established, 
we know almost nothing about the particle nature, except that it is,
to a high extent, electrically neutral.

One of the most appealing and natural CDM particle candidates is 
{\it weakly-interacting massive particle} (WIMP).  It is a coincidence
that if the dark matter is produced thermally in the early Universe,
the required annihilation cross section is right at the order of
weak interaction.  The relation between the relic density and
the thermal annihilation cross section can be given by the following 
simple formula \cite{hooper}
\begin{equation}
\label{rate}
\Omega_\chi h^2 \simeq \frac{ 0.1 \;{\rm pb} }{\langle \sigma v \rangle} \;,
\end{equation}
where $\langle \sigma v \rangle$ is the 
annihilation rate of the dark matter around the time of freeze-out.
Given the measured $\Omega_{\rm CDM} h^2$ the annihilation
rate is about $1$ pb or $10^{-26}\;{\rm cm}^3 \, {\rm s}^{-1}$.  
This is exactly
the size of the cross sections that one expects from a weak interaction
process and that would give a large to moderate production rate at the LHC.
In general, production of dark matter at the LHC would give rise to
a large missing energy.  Thus, the anticipated signature in the 
final state is high-$p_T$ jets or leptons plus a large missing energy.
Note that there could be non-thermal sources for the dark matter, such 
as decay from exotic relics like moduli fields, cosmic strings, etc. In 
such cases, the annihilation rate in Eq.~(\ref{rate}) can be larger than
the value quoted above.

The most studied dark matter candidate is perhaps the
neutralino of the supersymmetric models with $R$-parity conservation.
In this work, we study a different scenario.  The dark matter is
in a hidden sector and the only standard model (SM) particle that
it interacts with is the top quark. 
The top quark, having a mass so close to the electroweak symmetry 
breaking scale,
makes itself so unique among the fermions.  It is perhaps one of the
best windows to probe the electroweak symmetry breaking.  
The dark matter, if it is a WIMP, is also closely related to electroweak
symmetry breaking.  The logic is that since both the top quark and the
WIMP are closely related to electroweak symmetry breaking, we argue
that the top quark may be the only window to probe the dark matter.
This is our motivation. 
\footnote
{A possibility of realizing such a scenario can be found
in Ref.~\cite{servant}
}
We  use an effective Lagrangian approach to
parameterize the interactions between the top quark and the dark
matter particle, without specifying the detailed communication 
between the top quark and the hidden sector.  One simple example
would be a hidden-sector gauge boson that can couple to the
top quark.  If it is heavy enough we can shrink the propagator
to a 4-fermion vertex.  One form of the 4-fermion interaction is
$(\bar t \gamma_\mu t)\, (\overline{\chi} \gamma^\mu \chi)$ for a 
vector-type interaction or 
$(\bar t \gamma_\mu \gamma^5 t)\, (\overline{\chi} \gamma^\mu \gamma^5\chi)$ 
for an axial-vector-type interaction.
We can estimate the size of the new interaction based on the 
fact that it is the only interaction that can thermalize the 
dark matter particle in the early Universe.  
The most interesting implication of the scenario is the collider signature.  
The final state consists of a top-quark pair and
a pair of dark matter particles, giving rise to a top-quark pair plus
a large missing energy.
On the other hand, we  anticipate that the spin-independent 
and spin-dependent cross 
sections in direct detection would be consistent with the current
limit.  This is easy to understand because the top content inside the
nucleon is so small that it hardly contributes to the DM-nucleon 
scattering. 
We will explicitly show that. 
In addition, the annihilation of the dark matter in the Galactic Halo
would give rise to positrons and antiprotons that can be observed by
antimatter search experiments, e.g., PAMELA and AMSII.  We use the
present data on positron and antiproton spectra from PAMELA to 
constrain the size of the effective interactions.

The organization of the paper is as follows.  In the next section, we describe 
the interaction between the top quark and the dark matter particle,
and estimate the size of the interaction based on the relic density.
In Sec. III, we calculate the spin-independent and spin-dependent cross
sections for direct detection.  
In Sec. IV, we calculate the positron and antiproton spectra due to the
DM annihilation in Galactic halo.
In Sec. V, we discuss the collider signature.  Finally, we
conclude in Sec. VI.

There are a few recent works \cite{cao,bai,tait} that assumed some form of
effective interactions between the dark matter and light quarks and 
studied the corresponding collider phenomenology. Fan {\it et al.} \cite{fan}
also wrote down the effective nonrelativistic interactions between the dark
matter and nuclei.

\section{Effective Interactions and  Relic Density}

Our simple model consists of the SM and a hidden sector, in which
there is a pair of Dirac/Majorana fermions and a gauge boson.  For 
some reasons this gauge boson couples this hidden fermion only 
to the top quark on the SM side.  If the mass of this gauge boson is
heavy enough, we can integrate it out.  
More generally, below the heavy mass scale $\Lambda$ the 
interaction between the top quark and the dark matter particle, denoted
by $\chi$, is given by
\begin{equation}
\label{eff}
{\cal L} = \frac{g_\chi^2}{\Lambda^2} \,
   \left ( \overline{\chi} \Gamma \chi \right )\;
           \left ( \bar{t} \Gamma t \right ) \;,
\end{equation}
where $\Gamma = \gamma^\mu$ for a vector gauge boson, $\Gamma =
\gamma^\mu \gamma^5$ for an axial-vector gauge boson, $\Gamma=1 \,
(\gamma^5) $ for a scalar (pseudoscalar) boson interaction, and 
$\Gamma = \sigma^{\mu\nu} (\gamma^5)$ with 
$\sigma^{\mu\nu} \equiv i (\gamma^\mu \gamma^\nu -\gamma^\nu \gamma^\mu ) /2$ for
a tensor (axial-tensor) interaction, and
$g_\chi$ is an effective coupling constant.
For Majorana fermions the $\Gamma= \gamma^\mu$ or $\sigma^{\mu\nu}$ type 
interaction is identically zero, and so for vector or tensor type interaction 
the fermion $\chi$ in Eq.(\ref{eff}) must be Dirac. Explicitly, we 
assume the dark matter candidate to be Dirac, but the results
are also applicable to Majorana dark matter.
With this interaction we can calculate the thermal averaged cross sections
and thus the relic density, the direct and
indirect detection rates, and also
the production cross section of $pp \to t \bar t + \chi \overline{\chi}$
at the LHC.

\begin{figure}[t!]
\centering
\includegraphics[width=4.5in]{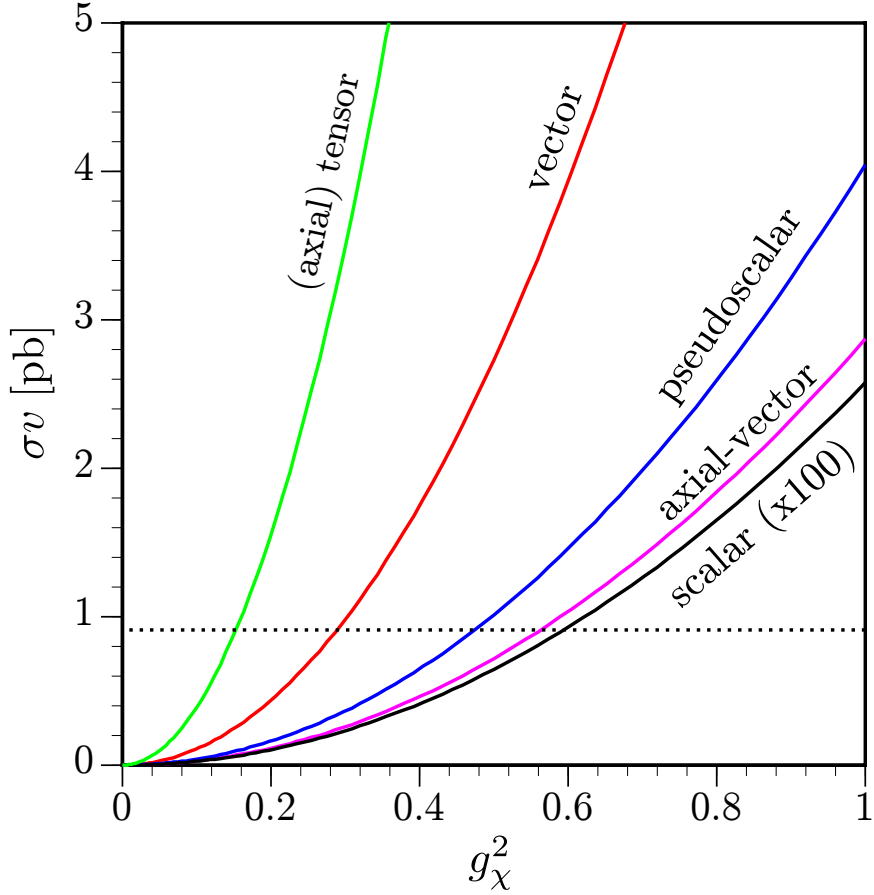}
\caption{\small \label{fig1}
The calculated $\sigma \, v$ versus $g_\chi^2$ for
the effective interaction $\frac{g_\chi^2}{\Lambda^2} \,
   \left ( \overline{\chi} \Gamma \chi \right )\;
           \left ( \bar{t} \Gamma t \right ) $ 
           of various Dirac structures $\Gamma$ with
$\Lambda = 1$ TeV, $m_\chi = 200 $ GeV, and $v \approx 0.3$.
}
\end{figure}

\begin{figure}[t!]
\centering
\includegraphics[width=4.5in]{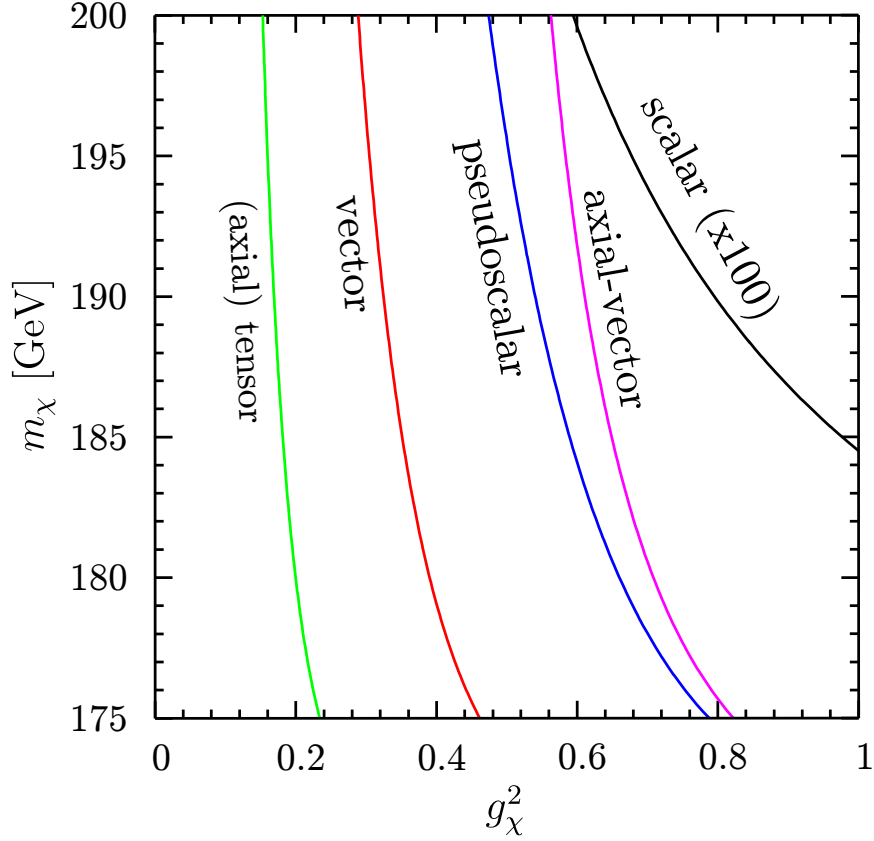}
\caption{\small \label {fig2}
Contours of $\sigma v = 0.91$ pb in the plane of 
$(g_\chi^2,\;m_\chi)$
for vector, axial-vector, pseudoscalar, scalar, tensor and 
axial-tensor interactions.  $\Lambda$ is set at 1 TeV.}
\end{figure}

We start with a vector gauge boson type interaction: $\Gamma = \gamma^\mu$.
The differential cross section 
for $\chi (p_1) \; \overline\chi (p_2) \to t (k_1) \;\bar  t (k_2)$, 
with the 4-momenta listed in the parentheses, is
\begin{equation}
\label{dsigma}
\frac{d\sigma}{dz}  = \frac{g_\chi^4}{\Lambda^4}\frac{N_C}{16\pi s} \,
\frac{\beta_t}{\beta_\chi}\, \left[ 
   u_m^2 + t_m^2 + 2 s ( m_\chi^2 + m_t^2)  \right ]
\end{equation}
where $N_C=3$ for the top quark color,
$\beta_{t,\chi} = ( 1- 4 m^2_{t,\chi}/s)^{1/2}$, 
$t_m = t - m_\chi^2 - m_t^2 = - s( 1- \beta_t \beta_\chi z)/2$,
$u_m = u - m_\chi^2 - m_t^2 = - s( 1 + \beta_t \beta_\chi z)/2$, 
$s=(p_1+p_2)^2$ is the
square of the center-of-mass energy, 
$t=(p_1 - k_1)^2$,
$u=(p_1- k_2)^2$,
and $z \equiv \cos\Theta$ with
$\Theta$ the scattering angle. The quantity $\sigma v$, where $v\approx
2 \beta_\chi$ in the non-relativistic limit, can be obtained by
integrating over the variable $z$ in Eq. (\ref{dsigma}). 
Instead of solving the Boltzmann equation,
we can naively estimate the size of the interaction by the
following equation
\begin{equation}
\Omega_\chi h^2 \simeq \frac{ 0.1\; {\rm pb}}{\langle \sigma v \rangle }\;.
\end{equation}
With the most recent WMAP result on dark matter density
$\Omega_{\rm CDM} h^2 = 0.1099 \pm 0.0062$ we obtain the size of $\sigma v$
\begin{equation}
\langle \sigma v \rangle \simeq 0.91  \; {\rm pb} \;.
\end{equation}
We show in Fig. \ref{fig1} $\sigma v$ versus the coefficient $g_\chi^2$
for a dark matter mass of 200 GeV.
The result shown is for $v\approx 0.3$ to approximate the velocity
of the dark matter particle at around the freeze-out time.
We can repeat the calculation with $\Gamma = 
\sigma^{\mu\nu}(\gamma^5), \; \gamma^\mu \gamma^5, \; \gamma^5,1$ 
for tensor (axial-tensor), axial-vector, pseudoscalar, and scalar 
type interactions. 
The results are
\begin{eqnarray}
\frac{d\sigma}{dz}  &=& 
\label{tensor-axialtensor}
\frac{g_\chi^4}{\Lambda^4}\frac{N_C}{4\pi s} \,
\frac{\beta_t}{\beta_\chi}\, \left[ 
  2 \, (t_m^2  + u_m^2 )+ 2 s (  m_t^2 +m_\chi^2 ) + 
8 m_t^2 m_\chi^2 - s^2 \right ]   \\
\frac{d\sigma}{dz}  &=& \frac{g_\chi^4}{\Lambda^4}\frac{N_C}{16\pi s} \,
\frac{\beta_t}{\beta_\chi}\, \left[ 
   t_m^2 + u_m^2 - 2 s ( m_t^2 + m_\chi^2 ) + 16 m_t^2 m_\chi^2 \right ] \\
\frac{d\sigma}{dz}  &=& \frac{g_\chi^4}{\Lambda^4}\frac{N_C}{32\pi} s 
  \frac{\beta_t}{\beta_\chi} \\
\frac{d\sigma}{dz}  &=& \frac{g_\chi^4}{\Lambda^4}\frac{N_C}{32\pi} s 
  \beta_\chi \beta_t^3 
\end{eqnarray}
for $\Gamma = \sigma^{\mu\nu}(\gamma^5), \; \gamma^\mu \gamma^5,\; 
\gamma^5,\; 1$, respectively.
We note that the axial-tensor case has the expression as in the
tensor one given by Eq.(\ref{tensor-axialtensor}).
The results are shown in Fig. \ref{fig1} as well. 
We can see that the tensor-type interaction gives the largest cross section,
followed by vector, pseudoscalar, and axial-vector.
These four types of interactions 
require $g_\chi^2$ falling into the range of  $0.2 -0.6$
which is about the size of weak-scale interaction.
On the other hand, the scalar-type interaction always gives a very small
annihilation cross section for a similar range of $g_\chi^2$, which
is in danger of over-closing the Universe.

In Fig. \ref{fig2}, we show the contour of the cross section for the various 
types of interactions
as a function of 
 $g^2_\chi$ and $m_\chi$ as allowed by the WMAP result. 

\section{Direct Detection}

Recently, the CDMSII finalized their search in Ref. \cite{cdms}.
When they opened the black box in their blind analysis,
they found two candidate events, which are
consistent with background fluctuation at a probability level of
about 23\%.  Nevertheless, the signal is not conclusive.  
The CDMS then improves 
upon the upper limit on the 
spin-independent (SI) cross section $\sigma^{\rm SI}_{\chi N}$ to
$3.8 \times 10^{-44} \; {\rm cm}^2$ for $m_\chi \approx 70 $ GeV.
The XENON100 Collaboration \cite{xenon100}
also recently announced their newest result.
Although XENON100 has the best sensitivity in the lower mass range, 
the CDMSII limit is currently still the best in the world for dark
matter mass larger than about 100 GeV.  
We will adopt a limit of order $4-10 \times 10^{-44}\; {\rm cm}^2$ 
for dark matter mass of $200-500$ GeV.
In the following, we will check if the spin-independent cross section
generated by the 4-fermion interactions is consistent with  
the new limit.

Spin-independent cross sections can arise from the scalar-type and
vector-type interactions between the DM and quarks. 
If the effective interactions between the dark matter particle 
and the quarks are given by
\begin{equation}
{\cal L } = \sum_{q=u,d,s,c,b,t} \{ \alpha_q^S \,\overline{\chi} \chi \, 
\bar q q +
   \alpha_q^V \,\overline{\chi} \gamma^\mu \chi \, \bar q \gamma_\mu q \} \;\; ,
\end{equation}
then the spin-independent cross section between $\chi$ 
and each of the nucleon (taking the average between proton and neutron)
is given by
\begin{equation}
\label{111}
\sigma^{\rm SI}_{\chi N} = \frac{4 \mu_{\chi N}^2}{\pi}\; \left (
 \left|G^N_s \right |^2 + \frac{|b_N|^2}{256} \right )\;,
\end{equation}
where $\mu_{\chi N} = m_\chi m_N / ( m_\chi + m_N)$ is the reduced mass
between the dark matter particle and the nucleon $N$, and 
\begin{equation}
G^N_s = \sum_{q=u,d,s,c,b,t} \langle N | \bar q q | N \rangle \;
  \alpha^S_q \;,
\end{equation}
where $\langle N | \bar q q | N \rangle$ denotes the various 
nucleon matrix elements.
The expression for $b_N$ of a {\em whole nucleus} $(A,Z)$ is
$b_N \equiv \alpha_u^V (A+Z) + \alpha^V_d (2A-Z)$, we take the 
average between proton and neutron (assume the number of protons is 
about the same as that of neutrons in the nuclei) and thus obtain the expression
for a single nucleon 
\begin{equation}
b_N = \frac{3}{2} \left( \alpha_u^V + \alpha_d^V \right ) \;\; .
\end{equation}
Nevertheless, the contributions to $b_N$ come from valence quarks only.
Therefore, in our scenario the only contribution to the SI cross section
comes from the top quark, 
thus only $\alpha_t^S $ is nonzero in the expression of $G^N_s$, which
is then given by
\begin{equation}
G^N_s =  \langle N | \bar t t | N \rangle \;
\left ( \frac{g^2_{\chi}}{\Lambda^2} \right  ) \;\;, 
\end{equation}
where $\langle N | \bar t t | N \rangle = f^N_{Tt} (m_N/m_t)$.
\footnote
{Equivalently, the top content inside the nucleon can be replaced by 
the gluon content with $f^N_{Tq}$ replaced by $\frac{2}{27} f^N_{Tg}$ 
\cite{hooper}. Numerically, they are very close to each other.}
The default value of the parameters $f^N_{Tt}$ used, 
e.g. in DarkSUSY \cite{darksusy}, is 
\[
f^p_{Tt} =  0.0595\;, \qquad  f^n_{Tt} =  0.0592 \;.
\]
Taking the average between proton and neutron the value of $G^N_s$ is
\begin{equation}
G^N_s \simeq  \frac{f^N_{Tt} m_N}{m_t} \left( \frac{g^2_{\chi}}{\Lambda^2} 
 \right ) \;.
\end{equation}
For $m_\chi \sim O(100)$ GeV, $\mu_{\chi N} \approx m_N$.  The 
spin-independent cross section is
\begin{equation}
\sigma^{\rm SI}_{\chi N} \approx \frac{ 4 \mu^2_{\chi N}}{\pi} \left(
      \frac{f^N_{Tt} m_N}{m_t} \right )^2 \, 
\left( \frac{g^2_{\chi}}{\Lambda^2} \right )^2 \; .
\end{equation}

\begin{figure}[t!]
\centering
\includegraphics[width=4in]{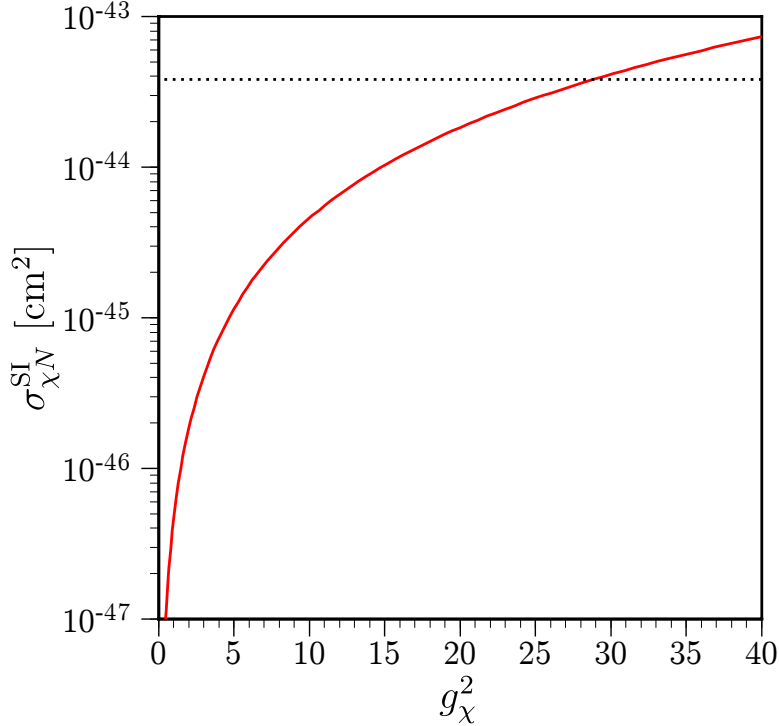}
\caption{\small \label{si}
Spin-independent cross sections for the vector type interaction versus
$g_\chi^2$.}
\end{figure}

We show in Fig.~\ref{si} the spin-independent cross section versus
$g_{\chi}^2$.  
Note that the axial-vector interactions contributes to spin-dependent
cross sections.
Since the constraint from spin-dependent cross sections is a few orders
of magnitude weaker than that from spin-independent cross sections,
we simply focus the spin-independent one to obtain the meaningful range
of $g_\chi^2$ and $\Lambda$.  We found that the limit on 
spin-independent cross section of the order of $10^{-44}\; {\rm cm}^{2}$
allows $g_\chi^2$ as large as $30$
 for $\Lambda = 1$ TeV. Note that for
a strongly coupled theory, one can have $g^2 = (4 \pi)^2$.  
Such a large $g_\chi^2$ is allowed by spin-independent cross section constraint 
as well as by the WMAP relic density constraint.
\footnote{
When the annihilation cross section is larger than that required by
thermal production, the resulting relic density from thermal production
is just too low.  However, there could be some other non-thermal
sources, such as decay from heavier fields. 
}
However, one must be cautious that for such a large effective 
coupling constant, perturbative calculation becomes less reliable.

We next turn to the indirect detection of the dark matter, which
then gives the strongest constraint on the present scenario.

\section{Indirect Detection}

Another important method to detect the dark matter is by measuring 
its annihilation products in Galactic halo. Current experiments can
detect the positron, antiproton, gamma ray, and deuterium from
dark matter annihilation.   The Milky Way Halo may contain
clumps of dark matter, from where the annihilation of dark matter particles
may give rise to large enough signals, such as positron and antiproton,
that can be identified by a number of antimatter search experiments. 
The most recent ones come from PAMELA \cite{pamela-e,pamela-p}, 
which showed a spectacular
rise in the positron spectrum but an expected spectrum for antiproton. 
It may be due to nearby pulsars or dark matter annihilation or decays.
If it is really due to dark matter annihilation, the dark matter
would have very strange properties, because it only gives positrons in 
the final products but not antiproton.  Here we adopt a conservative 
approach.  We use the observed antiproton and positron spectra as constraints
on the annihilation products in $\chi\bar \chi$ annihilation.

We first consider the positron coming from the process
\begin{equation}
 \chi \bar \chi \to t \bar t \to ( b W^+) (\bar b W^-) \to (b e^+ \nu_e) + X
\end{equation}
in which the most energetic $e^+$ comes from the $W^+$ decay.  There
are also positrons coming off in the subsequent decays of 
$b,\bar b$, $\tau^+$, or $\mu^+$, but these positrons are in general softer
than those coming directly from the $W^+$ decay.  For a first order 
estimate of the size of the coupling $g_\chi^2$ in Eq.~(\ref{eff}) we
only include the positrons coming directly from the $W^+$ decay.

\begin{figure}[t!]
\centering
\includegraphics[width=6in]{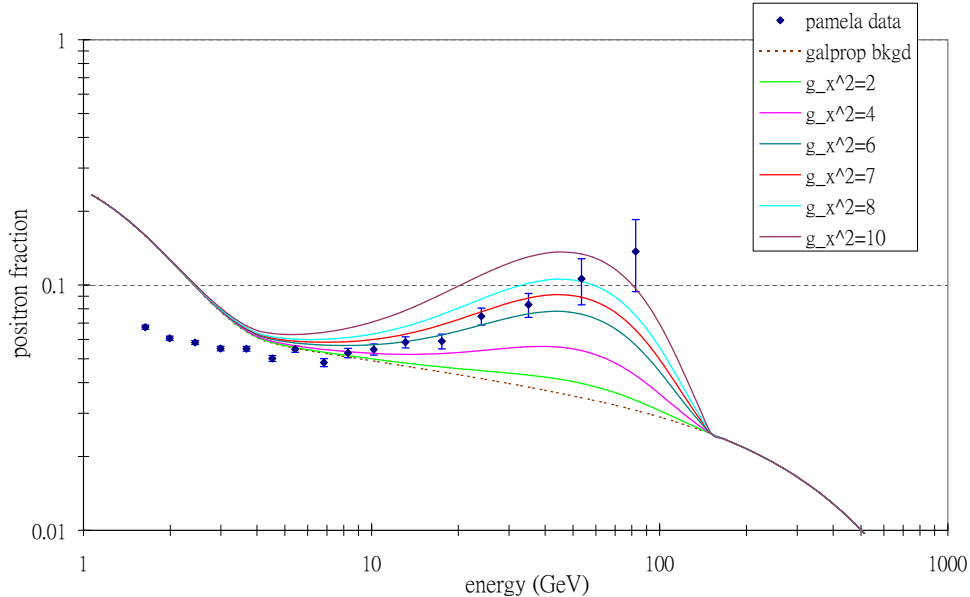}
\caption{\small \label{e+}
Spectrum for the positron fraction predicted for the vector type
interactions for various $g_\chi^2$.  PAMELA data are shown.}
\end{figure}

The expression for annihilation has already been given in Eq.~(\ref{dsigma}),
but now with a present time velocity $v\approx 10^{-3}$. 
The positron flux observed at the Earth is given by
\begin{equation}
 \Phi_{e^+} (E) = \frac{ v_{e^+} } { 4 \pi} \, f_{e^+} (E) \;,
 \label{semiflux} 
\end{equation}
with $v_{e^+}$ close to the velocity of light $c$.
The function 
$f_{e^+} (E)$ satisfies the diffusion equation of 
\begin{equation}
\frac{\partial f}{\partial t} - K(E)  \nabla^2 f 
- \frac{\partial}{\partial E} \left( b(E) f \right ) = Q \;,
\end{equation}
where the diffusion coefficient is
$K(E) = K_0(E/{\rm GeV})^\delta$ and 
the energy loss coefficient is 
$b(E) = E^2/ ({\rm GeV} \times \tau_E)$
with $\tau_E = 10^{16}$ sec. 
The source term $Q$ due to the annihilation is 
\begin{equation}
Q_{\rm ann} = \eta \left( \frac{\rho_{\rm CDM} }{M_{\rm CDM}} \right )^2 
\, \sum \langle \sigma v \rangle_{e^+} \, \frac{d N_{e^+}}{ d E_{e^+} } \;,
\end{equation}
where $\eta = 1/2 \; (1/4)$ for (non-)identical DM particle in the
initial state.
The summation is over all possible channels that can produce positrons in 
the final state, and $dN_{e^+}/dE_{e^+}$ denotes the spectrum of the positron
energy per annihilation in that particular channel. 
In our analysis, we employ the vector-type interaction for Dirac fermions
and thus the source term is given by
\begin{equation}
Q_{\rm ann} = \frac{1}{4} \left( 
\frac{\rho_{\rm CDM} }{M_{\rm CDM}} \right )^2 
\,   \langle \sigma v \rangle_{\chi\bar \chi\to t\bar t} \, \frac{d N_{e^+}}
  { d E_{e^+} } \;,
\end{equation}
where the normalization of $N_{e^+}$ is
\begin{equation}
\int \frac{d N_{e^+} }{d x} dx = B(t \to b W^+ \to b e^+ \nu_e) \;.
\end{equation}
We then put the source term into GALPROP\,\cite{GALPROP} 
to solve the diffusion equation.  In Fig.~\ref{e+} we show the
predicted energy spectrum for the positron fraction for various 
values of $g_\chi^2$.  With a visual inspection the $g_\chi^2 \alt 8$
is allowed by the spectrum.

\begin{figure}[t!]
\centering
\includegraphics[width=6in]{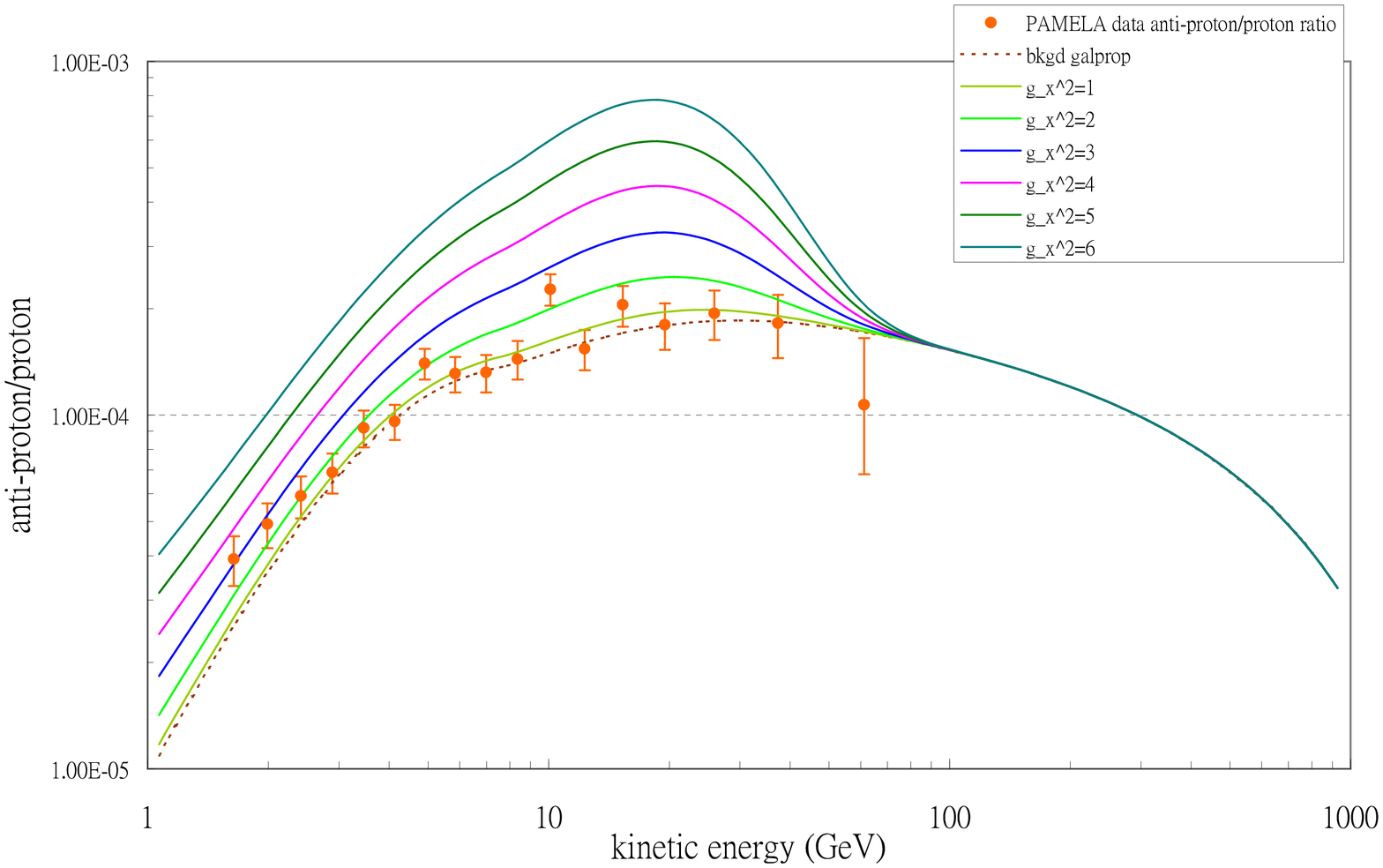}
\caption{\small \label{pbar}
Spectrum for the antiproton fraction predicted for the vector type
interactions for various $g_\chi^2$.  PAMELA data are shown.}
\end{figure}

Next we turn to the antiproton fraction as it was also measured by PAMELA.
Similarly, the antiproton flux can be obtained by solving the 
diffusion equation with the corresponding terms and the 
appropriate source term for the input antiproton spectrum:
\begin{equation}
Q_{\rm ann} = \eta \left( \frac{\rho_{\rm CDM} }{M_{\rm CDM}} \right )^2 
\, \sum \langle \sigma v \rangle_{\bar p} \, \frac{d N_{\bar p}}{ d T_{\bar p} }
 \;,
\end{equation}
where $\eta =1/2\;(1/4) $ for (non-)identical initial state, and $T_{\bar p}$ is 
the kinetic energy of the antiproton
which is conventionally used instead of the total energy. 
We again solve the diffusion equation using GALPROP\,\cite{GALPROP}.

In our case, the dominant contribution comes from
\begin{equation}
\chi \bar \chi \to t \bar t \to (b W^+) (\bar b W^-) \to ( b q \bar q')
(\bar b q \bar q') \to \bar p + X \;.
\end{equation}
In the last step, all the $b\,\bar b, q, \bar q'$ have probabilities 
fragmenting into $\bar p$.  We adopt a publicly available
code\,\cite{kniehl} to calculate the fragmentation function $D_{q\to h}(z)$ 
for any quark $q$ into hadrons $h$, e.g., $p,\bar p, \pi$.  
The fragmentation
function is then convoluted with energy spectrum $d N/ dE$ 
of the light quark to obtain the energy spectrum of the antiproton
$d N /d E_{\bar p}$.  
The source term $dN /d T_{\bar p}$ is then 
implemented into
GALPROP to calculate the propagation from the halo to the Earth.
We display the energy spectrum for the antiproton fraction in 
Fig.~\ref{pbar}.  It is easy to see that $g_\chi^2$ is constrained to
be 
\begin{equation}
 g^2_\chi \alt 4-5 \;.
\end{equation}
We will use this allowed range to estimate what we would expect from the
LHC.

\section{Collider Signature}

\begin{figure}[t!]
\centering
\includegraphics[width=3in,clip]{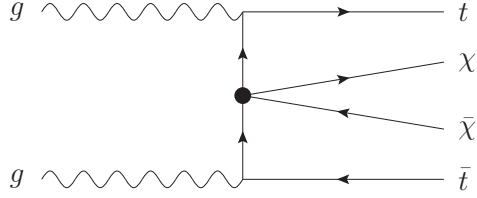}
\caption{\label{feyn} \small 
A contributing Feynman diagram for the subprocess $gg \to t \bar t + \chi
\overline{\chi}$. The other two diagrams can be obtained by attaching
the black dot to the $t$ and $\bar t$ leg, respectively.}
\end{figure}

Collider signatures are perhaps the most interesting part
of the scenario -- $t\bar t$ pair plus large missing energy. 
We first calculate
using the effective 4-fermion interaction with $\Gamma = \gamma^\mu$ 
the production cross
section for $pp \to t \bar t + \chi \bar \chi$.  There are two contributing
subprocesses for $t\bar t$ production at the LHC:
\begin{equation}
q\bar q \to t \bar t\;, \qquad \qquad gg \to t \bar t \;,
\end{equation}
on which we can attach one 4-fermion interaction vertex to each fermion 
leg including internal fermion line to further produce a $\chi \overline{\chi}$
pair. A typical Feynman diagram is shown in Fig. \ref{feyn}.
We employ MADGRAPH \cite{madgraph} to calculate the signal and 
background cross sections.

\begin{figure}[t!]
\centering
\includegraphics[width=4in,clip]{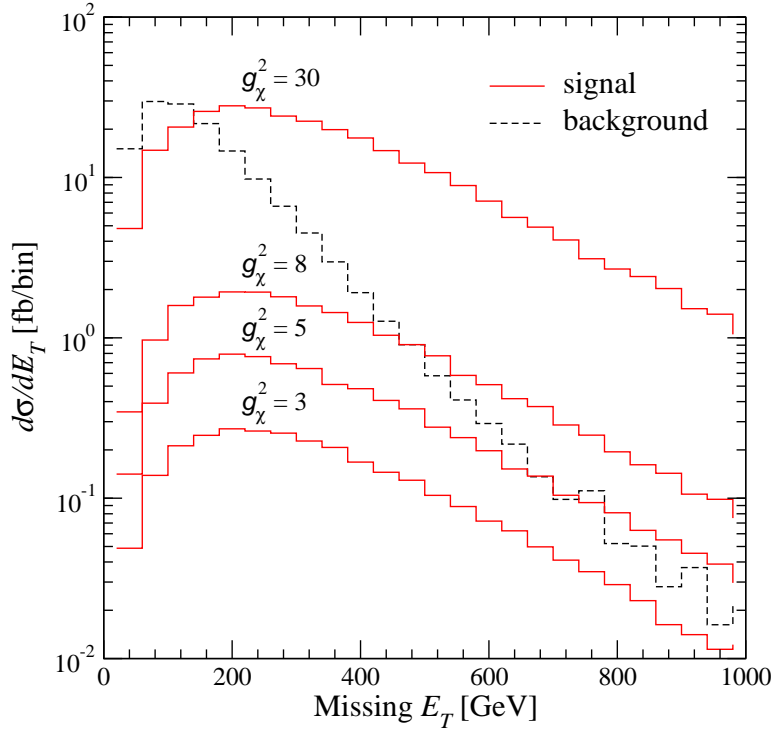}
\caption{\small \label{fig7}
Missing transverse energy $\not\!{E}_T$
distributions for the signal $ pp \to t\bar t + \chi \bar \chi$ and 
the background $pp\to t\bar t Z$ for $g_\chi^2=3-30$ with $m_\chi=200$ GeV.}
\end{figure}

\begin{figure}[t!]
\centering
\includegraphics[width=6in,clip]{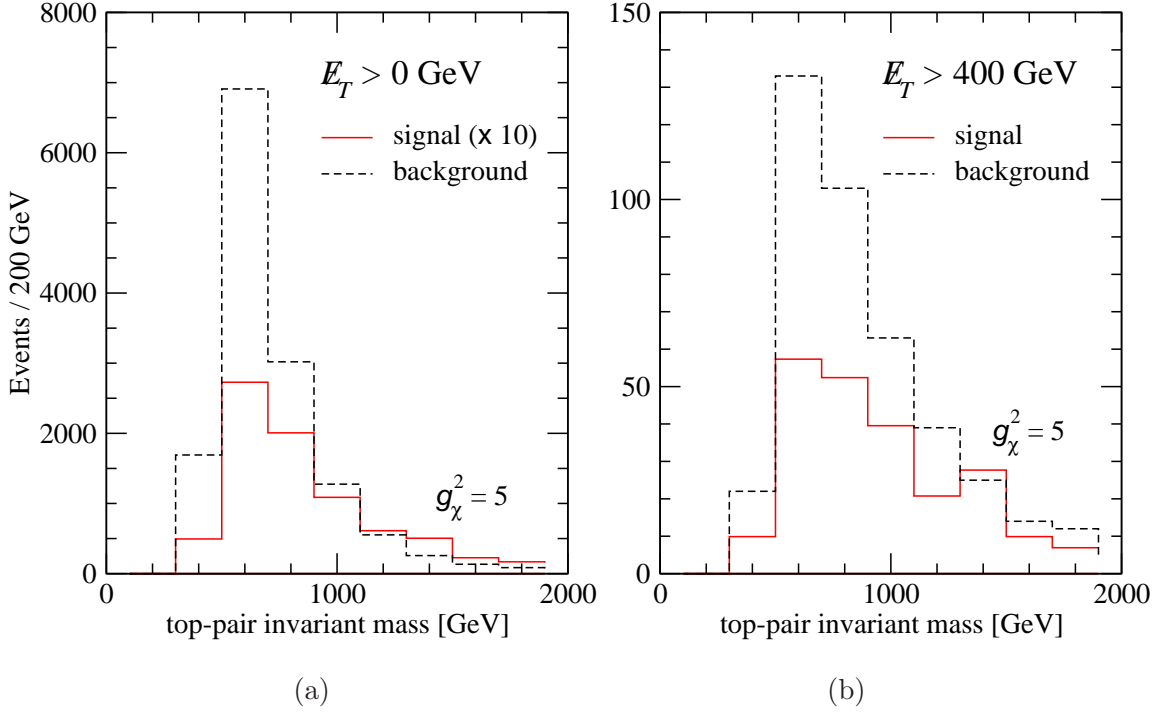}

(a)  \hspace{2.5in}                       (b) 
\caption{\small \label{fig8}
Event numbers for the invariant mass $t\bar t$ distributions for the signal 
$ pp \to t\bar t + \chi \bar \chi$ and the background $pp\to t\bar t Z$
(a) before and (b) after applying the missing transverse momentum 
cut of 400 GeV.
The assumed luminosity is 100 fb$^{-1}$, which corresponds to 240 signal events
and 420 background events after the cut.}
\end{figure}

The irreducible background is $t\bar t + Z \to t\bar t \nu\bar \nu$.
Before applying any cuts we calculate the signal cross section versus
background cross section: 
8.2 fb (for $g_\chi^2=5$) to 140 fb,
in which we have chosen scale $Q = (2m_t + 2m_\chi)/2$ in the running
coupling constant and the parton distribution functions for the signal, 
while $Q= (2m_t + m_Z)/2$ for the background.
We first compare the missing $E_T$ distribution between the dark 
matter signal and the $t\bar t Z$ background, shown in Fig. \ref{fig7}.
It is clear that the signal has a harder missing $E_T$ spectrum than
the background. 
This plot suggests a cut as large as 400 GeV in the missing 
transverse energy can substantially reduces
the background to a level similar to the signal.
The cross sections in fb for the signal and the background
using various cuts on the missing energy are shown in Table~\ref{table}.
The background can indeed be cut down to the level as the signal with
a missing energy cut of 400 GeV.  Note that the signal cross section
scales as $g_\chi^4$.  The significance of the signal $S/\sqrt{B}$ for
an integrated luminosity of 100 fb$^{-1}$ stays around $11$ with a cut 
of $300-500$ GeV.  Since the significance scales as $\sqrt{\cal L}$,  with a 
reduced luminosity of 30 fb$^{-1}$ the significance is still as large as
$6$. 

We then compare the $t\bar t$ invariant mass distribution 
between the dark matter signal and the $t\bar t Z$ background
before and after applying the missing $E_T$ cuts, 
shown in Fig.~\ref{fig8}.
In Table~\ref{table2}, we also show the signal cross sections
and the significance for axial-vector, pseudoscalar, and scalar interactions
\footnote
{
Note that the tensor interaction is not present in current version of 
MADGRAPH \cite{madgraph}.}
in decreasing order.  Note that the cross section at the LHC for 
scalar interaction is not severely suppressed, in sharp 
contrast to the annihilation cross section calculated in Sec. II.

\begin{table}[t!]
\centering
\caption{\small \label{table}
Cross sections in fb for the signal $pp \to t\bar t + \chi\bar\chi$
and the background $pp\to t\bar t Z \to t\bar t + \nu\bar\nu$ at the LHC.
We used $g_\chi^2 = 5$ for illustration. The signal cross section scales as
$g_\chi^4$. The significance $S/\sqrt{B}$ is calculated with an integrated
luminosity of 100 (30) fb$^{-1}$. }
\medskip
\begin{ruledtabular}
\begin{tabular}{lllll}
$\not\!{E}_T > $   &  $pp\to t\bar t + \chi\bar \chi$  & 
 $p\to t\bar t Z \to  t\bar t \nu \bar \nu$  & $S/B$ & 
  $S/\sqrt{B}$ (100 (30)  fb$^{-1}$)\\
\hline
$0$ GeV   & $8.2$ & $140.3$ & $0.06$  & $6.9\;(3.8) $ \\
$300$ GeV & $3.6$ & $10.7$ & $0.34$  & $11.0\;(6.0)$ \\
$400$ GeV & $2.4$ & $4.2$  & $0.57$  & $11.8\;(6.4) $ \\
$500$ GeV & $1.5$ & $1.9$  & $0.78$  & $10.6\;(5.9) $ 
\end{tabular}
\end{ruledtabular}
\end{table}

\begin{table}[t!]
\centering
\caption{\small \label{table2}
Cross sections in fb for the signal $pp \to t\bar t + \chi\bar\chi$
for vector, axial-vector, pseudoscalar, and scalar interactions at the LHC.
We have imposed the $\not\!\!{E}_T > 400$ GeV cut.  The $S/B$ and $S/\sqrt{B}$
are shown. The background is from Table I. 
The significance $S/\sqrt{B}$ is calculated with an integrated
luminosity of 100 (30) fb$^{-1}$.
}
\medskip
\begin{ruledtabular}
\begin{tabular}{llll}
       & Signal cross section (fb)  &  $S/B$   &  $S/\sqrt{B}$ \\
\hline
Vector & $2.4$ &  $ 0.57$ & $11.8\;(6.4)$ \\
Axial-vector  & $1.9$ & $0.45$ & $9.3\;(5.1)$ \\
Pseudoscalar  & $0.82$ & $0.20$ & $4.0\;(2.2)$ \\
Scalar        & $0.55$ & $0.13$ & $2.7\;(1.5)$ \\
\hline
\end{tabular}
\end{ruledtabular}
\end{table}

\section{Conclusions}

In this paper we have studied an interesting scenario
where the dark matter couples exclusively to the top quark using
an effective field theory approach, with the intuition that
both the top quark and the dark matter may be closely related to
electroweak symmetry breaking.
We did not specify any
particular connector linking the SM sector and the invisible dark
matter sector, except that this connector sector was taken to be
heavy, probably at the TeV scale. Integrating out the heavy
connector sector may give rise to effective 4-fermion interactions
of tensor, axial-tensor, vector, axial-vector, pseudoscalar, or 
scalar types between the
dark matter and the top quark. We studied the constraints of these
effective couplings from WMAP as well as from the direct and indirect
detection of dark matter at CDMSII and PAMELA, respectively. 

If we require all the dark matter in the Universe
comes from the thermal equilibrium,
the coupling $g_\chi^2 \approx 0.3 - 0.6$.  However, if we just require
that the dark matter does not overclose the Universe the $g_\chi^2$ can be  
much larger.  Since only the top quark inside the nucleon contributes
to direct detection cross section, 
the coupling $g_\chi^2$ can be as large as $40$.
On the other hand, the strongest constraint comes from the 
positron and antiproton fraction spectra.  
The PAMELA antiproton spectrum constrains the coupling to be 
$g_\chi^2 \alt 4-5$.   

This model can be tested at colliders with a very distinct
signature, namely, $t\bar t$ plus missing energies.  The top quark and
antiquark would mostly have high $p_T$ and boosted.  The detection
of such boosted top quarks has attracted some recent studies that
it can be sufficiently distinguished from the background \cite{boost}.
Our results suggested that this interesting scenario can be testable at
the LHC.

\section*{Acknowledgments}
The work was supported in parts by the National Science Council of
Taiwan under Grant Nos. 96-2628-M-007-002-MY3, 99-2112-M-007-005-MY3, and
98-2112-M-001-014-MY3, the NCTS, and the
WCU program through the KOSEF funded by the MEST (R31-2008-000-10057-0).


\end{document}